\documentclass[superscriptaddress,showpacs,amssymb,10pt,reprint,aps,prd,longbibliography,nofootinbib,floatfix]{revtex4-1}

\usepackage{geometry}
\usepackage{graphicx,epsfig,amssymb}
\usepackage{amsmath,amsfonts, times}
\usepackage{bm}

\usepackage{graphicx}
\usepackage{subcaption}
\usepackage{float}

\usepackage{epstopdf}
\usepackage[linktocpage,colorlinks]{hyperref}
\usepackage[usenames]{color}

\usepackage{soul}
\usepackage[utf8x]{inputenc}
\usepackage[usenames]{color}
\usepackage{booktabs}
\usepackage{threeparttable}
\newcommand{\beq}{\begin{equation}}
	\newcommand{\eeq}{\end{equation}}
\newcommand{\bqn}{\begin{eqnarray}}
	\newcommand{\eqn}{\end{eqnarray}}
\begin{document}
	\title{Phase-Dependent Photon Emission Rates in Quantum Gravity-Induced Entangled States}
	\author{Chi Zhang}
	\email{zhangchi3244@gmail.com; Corresponding author}
	\affiliation{Department of Physics, Zhejiang Ocean University, Zhejiang, 316022, China}

\begin{abstract}
Quantum entanglement, as one of the fundamental concepts in quantum mechanics, has garnered significant attention over the past few decades for its extraordinary nonlocality. With the advancement of quantum technology, quantum entanglement holds promising application for exploring  fundamental physical theories. The experimental scheme of Quantum Gravity Induced Entanglement of Masses (QGEM) was proposed to investigate the quantum effects of gravity based on the Local Operations and Classical Communication (LOCC) theory. In this study, we analyze the quantum properties of the entangled final states generated in the QGEM scheme. Our findings reveal that the photon emission rates (transition rates) are closely related to the degree of entanglement. Specifically, the transition rate decreases as the degree of entanglement increases when the distance between particle pairs is small, then it gradually approaches an asymptotic value that is independent of entanglement as the distance increases. We then discuss the possibility of using photon emission rates to detect quantum entanglement with these results.
\end{abstract}
\date{\today}
\maketitle
\section{Introduction}
Quantum technology has undergone significant advancements in recent decades, enabling broader application in both theoretical exploration and industrial production. Among its fascinating phenomena, quantum entanglement \cite{Horodecki:2009zz}—characterized by unique nonlocal correlations—has emerged as a valuable resource with extensive research. For example, there have been considerations about entanglement-based tomography to probe new macroscopic forces \cite{Barker:2022mdz} and the highest-energy observation of entanglement in top–antitop quark events produced at the LHC was reported in \cite{ATLAS:2023fsd}. Quantum entanglement has also been widely applied in quantum computing and quantum communication \cite{Horodecki:2009zz,Braunstein:2005zz,r9,r10,r11} and in optical imaging by means of two-photon quantum entanglement \cite{r13}, etc.

The development of quantum technology has also inspired a series of table-top gravity related experiments. The theory of classical general relativity, established by Albert Einstein in 1916, has been successfully verified through numerous practical applications. However, this classical theory of gravity appears incompatible with the quantum mechanics. The debate over whether the gravitational field is a quantum field has lasted for a long time and experimental validation is required to reach a conclusive answer. Recently, the experimental scheme Quantum Gravity Inducing Entanglement of Masses (QGEM) proposed by Bose and Marletto et al. in \cite{r3,Marletto:2017kzi} utilized quantum entanglement to detect quantum gravity effects. Later on, people have developed more experimental ideas based on QGEM to enhance its feasibility and completeness \cite{r7,r19,r17,r18}. However, all the final states in these QGEM experiments can be represented as a single common state from which we could deduce quantum gravity. If the gravitational field can indeed act as a quantum communication channel and turn the initial direct product state into an entangled final state, then according to the Local Operations and Classical Communication (LOCC) theorem in quantum information theory, we can infer the gravitational field acts as a quantum communication channel and the quantum nature of the gravitational field.

The QGEM proposal provides an important framework for experimental investigations into the quantum aspects of Newtonian gravity. However, there are many debates about whether it offers conclusive evidence for the quantization of the gravitational field or the existence of gravitons. Several authors hold the view that it only focuses on Newtonian gravity, which lacks dynamical degrees of freedom \cite{Kaku:2024lgs,Carney:2021vvt,Martin-Martinez:2022uio}. Marchese et al. show that one can generate the same amount of entanglement in this setup by using classical time evolution given by Newton's laws of motion \cite{Marchese:2024zfu}. Fragkos et al. point out that even within relativistic physics, entanglement generation can equally be described in terms of mediators or in terms of non-local processes so it is not possible to draw conclusions about gravitational field \cite{Fragkos:2022tbm}. These shortcomings and controversies of QGEM deserve further study in the future.

The transition rate of entangled particle pair has been investigated in \cite{r8} to explore the feasibility of detecting the Unruh effect. Arias et al. \cite{Arias:2015moa} studied the spontaneous transition rates of two maximally entangled two-level atoms under vacuum fluctuations, both in the presence and absence of a perfect reflecting boundary. Barman et al. \cite{Barman:2022utm} examined the transition processes of two entangled particle detectors moving along circular orbits. Blaauboer analytically solved the time evolution of entangled electron spin pairs in a double quantum dot nanostructure \cite{r31}. However, research on the transition rates of entangled states of distinguishable particles generated via the QGEM mechanism remains unexplored.

In this paper, we focus on the entangled quantum states induced by the QGEM mechanism, particularly the effect of entanglement on particle spontaneous transition rates and corresponding photon emission rates. Through first-principles calculations, we demonstrate how entanglement suppresses or promotes the particle pair transitions, shedding light on entanglement dynamics and offering guidance for related experimental designs.

\section{Photon emission rate for different entanglement phases}
In the QGEM experiment, as a consequence of the gravitational interaction between the superposition branches of the two particles, quantum correlations between the two particles do arise as the formation of the entanglement phase $\delta \phi$ \cite{r3}. The positive spin direction is denoted as $\uparrow$ and negative spin direction as $\downarrow$. The entangled final state formed by the QGEM setup after a certain period of evolution is:
\beq\label{1}
\left| {\Phi} \right\rangle  = \frac{1}{2}\left( {\left| { \uparrow  \uparrow } \right\rangle  + \left| { \uparrow  \downarrow } \right\rangle  + {e^{i\delta \phi }}\left| { \downarrow  \uparrow } \right\rangle  + \left| { \downarrow  \downarrow } \right\rangle } \right).
\eeq
Note that in \cite{r3}, there are two superposition branches both having relative phases. However, for simplicity, we have only focused on the dominant entanglement phase in the branch with smallest spatial separation and ignored the smaller phases.

The entangled state is the overall state of particles 1 and 2, and the properties it describes are not the simple sum of the properties of the two particles. The inseparability of equation \eqref{1}, i.e. entanglement, is fully reflected in the entanglement phase ${\delta \phi }$, therefore ${\delta \phi }$ directly determines many properties of the entangled final state.

According to \cite{r3}, the Entanglement Witness(Wit) was selected as the experiment indicator to detect entanglement formation:
\beq\label{17}
 Wit = \left| {\left\langle {\sigma {1_x} \otimes \sigma {2_z}} \right\rangle  + \left\langle {\sigma {1_y} \otimes \sigma {2_y}} \right\rangle } \right|.
\eeq
When it's greater than 1, we can infer that there is entanglement between the two particles. 

While the entanglement witness is typically used to detect entanglement in experiment, we attempt to propose an alternative approach. By examining the transition rates of entangled particle pairs from higher energy level to lower energy level with photons emitted, we may infer the presence and degree of entanglement. The photon emission rate may serve as an indirect evidence of entanglement.

When an energy difference exists between particle's different magnetic dipole moment orientations, which may originate from the particle's own structure or an externally applied uniform magnetic field, the initial state with a higher energy level will emit a photon to spontaneously transition to the low energy state. Assume that the eigenstate of ${\sigma _z}$ with eigenvalue +1, denoted as ${\left|  \uparrow  \right\rangle }$, possesses a higher energy level than the eigenstate of ${\sigma _z}$ with eigenvalue -1, denoted as ${\left|  \downarrow  \right\rangle }$. According to the first-order perturbation theory, the final state with an emitted photon after the transition is:
\begin{widetext}
\beq\label{5}
\begin{split}
\left| {\Omega \left( t \right)} \right\rangle  = ( {a_1}\left( t \right)\left| { \uparrow  \uparrow } \right\rangle  + {a_2}\left( t \right)\left| { \uparrow  \downarrow } \right\rangle  + {a_3}\left( t \right)\left| { \downarrow  \uparrow } \right\rangle &+ {a_4}\left( t \right)\left| { \downarrow  \downarrow } \right\rangle ) \otimes \left| 0 \right\rangle \\
&+ \left( {a_5}\left( t \right)\left| { \uparrow  \uparrow } \right\rangle  + {a_6}\left( t \right)\left| { \uparrow  \downarrow } \right\rangle  + {a_7}\left( t \right)\left| { \downarrow  \uparrow } \right\rangle  + {a_8}\left( t \right)\left| { \downarrow  \downarrow } \right\rangle \right) \otimes \left| {photon} \right\rangle 
\end{split}
\eeq
\end{widetext}
where $\left| 0 \right\rangle$ denotes the vacuum state of photon field and the initial values of ${a_1}\left( t \right) \sim {a_4}\left( t \right)$ are precisely the coefficients of each superposition branch in Eq. \eqref{1}. In the following text, we take the following eight eigenstates as the basis:
\beq\label{23}
\begin{split}
&\left| 1 \right\rangle  = \left| { \uparrow  \uparrow } \right\rangle  \otimes \left| 0 \right\rangle ,\;\;\;\;\;\;\;\;\;\left| 2 \right\rangle  = \left| { \uparrow  \downarrow } \right\rangle  \otimes \left| 0 \right\rangle ,\\
&\left| 3 \right\rangle  = \left| { \downarrow  \uparrow } \right\rangle  \otimes \left| 0 \right\rangle ,\;\;\;\;\;\;\;\;\;\left| 4 \right\rangle  = \left| { \downarrow  \downarrow } \right\rangle  \otimes \left| 0 \right\rangle ,\\
&\left| 5 \right\rangle  = \left| { \uparrow  \uparrow } \right\rangle  \otimes \left| {photon} \right\rangle ,\left| 6 \right\rangle  = \left| { \uparrow  \downarrow } \right\rangle  \otimes \left| {photon} \right\rangle ,\\
&\left| 7 \right\rangle  = \left| { \downarrow  \uparrow } \right\rangle  \otimes \left| {photon} \right\rangle ,\left| 8 \right\rangle  = \left| { \downarrow  \downarrow } \right\rangle  \otimes \left| {photon} \right\rangle.
\end{split}
\eeq


The interaction Hamiltonian for magnetic dipole transitions is:
\beq\label{6}
\hat H' =  - \sum\limits_\eta  {{{\hat u}_e}\left( \eta  \right) \cdot \hat B\left( {{r_\eta }} \right)} 
\eeq
where ${\hat u\left( \eta  \right)}$, $\eta = 1,2$ are the magnetic dipole moments for particles 1 and 2 respectively. ${\hat B\left( {{r_\eta }} \right)}$ is the transverse magnetic field vector operator evaluated at the spatial position of 1 or 2 particles and may be expanded into the Fourier series \cite{r1}:
\beq\label{8}
\begin{array}{l}
	\hat B\left( r \right) = \sum\limits_k {\sum\limits_\lambda  i } \sqrt {\frac{\hbar }{{2{\varepsilon _0}{w_k}V}}} \vec k \times {{\vec e}_k}^{\left( \lambda  \right)}\left( {{{\hat a}_{k\lambda }}{e^{i\vec k \cdot \vec r}}} \right.\\[10pt]
	\;\;\;\;\;\;\;\;\left. { - \hat a_{k\lambda }^\dag {e^{ - i\vec k \cdot \vec r}}} \right)
\end{array}
\eeq
The quantization volume is V, ${w_k} = kc$ is the circular frequency and ${\vec k}$ is the wave vector. The state of polarization of the transverse magnetic field is defined by the index $\lambda $( = 1 or 2). 

Next we make use of first-order time-dependent perturbation theory to give photon emission rates from the initial state $\left| \Phi  \right\rangle $ to the final state $\left| \Omega  \right\rangle $\cite{r20}. We have Schrödinger equation in the interaction picture:
\beq\label{22}
i\hbar \frac{d}{{dt}}\left| {\tilde \Omega \left( t \right)} \right\rangle  = \tilde H'\left( t \right)\left| {\tilde \Omega \left( t \right)} \right\rangle,
\eeq
where
\beq
\left| {\tilde \Omega \left( t \right)} \right\rangle  = {e^{\frac{{i{{\hat H}_0}t}}{\hbar }}}\left| {\Omega \left( t \right)} \right\rangle, \tilde H'\left( t \right) = {e^{\frac{{i{{\hat H}_0}t}}{\hbar }}}\hat H'\left( t \right){e^{ - \frac{{i{{\hat H}_0}t}}{\hbar }}}.
\eeq
Expand Eq. \eqref{22} in terms of the eight basis in expression \eqref{23}, the coefficients of the superposition states in Eq. \eqref{5} evolve as:
\beq\label{18}
\begin{split}
	i\hbar {{\dot a}_m}(t) =\sum\limits_n {{e^{i\frac{{\Delta {E_{mn}}}}{\hbar }t}}} {{H'}_{mn}}{a_n}(t),\\
	m,n = 1,2,3,4,5,6,7,8 .
\end{split}
\eeq
Here ${{H'}_{mn}} = \left\langle m \right|\hat H'\left| n \right\rangle $, is the first-order infinitesimal quantity. Assume that the initial state $\left| \Phi  \right\rangle$ interacts with the vacuum electromagnetic field at $t=0$, the first-order iterative approximate solution of Eq. \eqref{18} in ${H'}$ for the superposition branch with photon emission is:
\beq\label{19}
\begin{split}
{a_m}(t) = \sum\limits_n {\frac{{1 - {e^{i\frac{{\Delta E_{mn}^{dipole} + {E_p}}}{\hbar }t}}}}{{\Delta E_{mn}^{dipole} + {E_p}}}} {{H'}_{mn}}{a_n}(0),\\
	n = 1,2,3,4; m = 5,6,7,8,
\end{split}\eeq
where ${\Delta E_{mn}^{dipole}}$ is the energy difference of dipole between basis $m$ an $n$, ${E_p}$ is the energy of photon emitted. 

Sum all the possible final states of emitted photon, the separate transition rate ${R_m}$ is given by:
\beq\label{9}
{R_m} = \frac{{d\left( {\int {{{\left| {{a_m}(t)} \right|}^2}\rho \left( {\vec k} \right){d^3}k} } \right)}}{{dt}},
\eeq
where $\rho \left( {\vec k} \right)$ is the density of photon in final states:
\beq\label{11}
\rho \left( {\vec k} \right){d^3}k = \frac{{V{k^2}}}{{{{\left( {2\pi } \right)}^3}}}d\Omega dk,\;\;\;\;\;\; k = \frac{{{E_p}}}{{\hbar c}}.
\eeq
Using rotating wave approximation, when t is large enough only the term with the same negative $\Delta {E_{mn}^{dipole}}$ in Eq. \eqref{19} contributes significantly to ${R_m}$ in Eq. \eqref{9}. When integrating, substitute the following formula into the integral expression \cite{r21}:
\beq\label{12}
\sum\limits_\lambda  {{e_\alpha }^{\left( \lambda  \right)}{e_\beta }^{\left( \lambda  \right)}}  = {\delta _{\alpha \beta }} - {{\hat k}_\alpha }{{\hat k}_\beta }. 
\eeq
In this paper, the subscripts $\alpha$ and $\beta$ stand for the three dimensions of the space, i.e. $\alpha,\beta = x,y,z$.
Then we get the photon emission rates expressions through Eq. \eqref{9}:
\begin{widetext}
\beq\label{14}
\begin{split}
	&{R_1} = 0\\
	&{R_2} = \frac{{\pi }}{2\hbar }\int {{{\left| {{{H'}_{61}}} \right|}^2}\rho \left( {{E_p}} \right)d\Omega }\\
	&{R_3} = \frac{{\pi }}{2\hbar }\int {{{\left| {{{H'}_{71}}} \right|}^2}\rho \left( {{E_p}} \right)d\Omega }\\
	&{R_4} = \frac{{\pi }}{2\hbar }\int {\left( {{{\left| {{{H'}_{82}}} \right|}^2} + {{\left| {{{H'}_{83}}} \right|}^2} + {{H'}_{82}}^*{{H'}_{83}} + {{H'}_{83}}^*{{H'}_{82}}} \right)\rho \left( {{E_p}} \right)d\Omega }, 
\end{split}
\eeq
\end{widetext}
where ${{E_p}}$ corresponds to the energy released by the flipping of a single magnetic dipole moment. The above photon emission rates consist with the well known fermi's golden rule \cite{r1,r21,r14}. Then the total photon emission rate is:
\begin{widetext}
\beq\label{20}
\begin{split}
R\left( {\phi ,k,d} \right) &= {R_1} + {R_2} + {R_3} + {R_4} \\
&= \frac{{{u_0}{k^3}}}{{8\pi \hbar }}\left( {\frac{2}{3}{\delta _{\alpha \beta }}\left( {\overline {u_{2\alpha }^{61}} {{\overline {u_{2\beta }^{61}} }^*} + \overline {u_{1\alpha }^{71}} {{\overline {u_{1\beta }^{71}} }^*} + \overline {u_{1\alpha }^{82}} {{\overline {u_{1\beta }^{82}} }^*} + \overline {u_{2\alpha }^{83}} {{\overline {u_{2\beta }^{83}} }^*}} \right)} \right.\left. { + 2{\rm{Re}}\left( {{\tau _{\alpha \beta }}\overline {u_{2\alpha }^{83}} {{\overline {u_{1\beta }^{82}} }^*}} \right)} \right).
\end{split}
\eeq
\end{widetext}

In the above \({u_0}\) is permeability of vacuum and k is the wave number of the emitted photon. ${u_{2\alpha }^{61}}$ is the expected value of magnetic dipole moment of particle 2 corresponding to the initial and final states in ${{{H'}_{61}}}$ and the rest are similar. And ${\tau _{\alpha \beta }}$ is \cite{r21}:
\beq\label{21}\begin{split}
&{\tau _{\alpha \beta }}\left( {kd} \right) \equiv {\alpha _{\alpha \beta }}\frac{{\sin kd}}{{kd}} + {\beta _{\alpha \beta }}\left( {\frac{{\cos kd}}{{{k^2}{d^2}}} - \frac{{\sin kd}}{{{k^3}{d^3}}}} \right),\\
&{\alpha _{\alpha \beta }} \equiv {\delta _{\alpha \beta }} - {{\hat d}_\alpha }{{\hat d}_\beta },\\
&{\beta _{\alpha \beta }} \equiv {\delta _{\alpha \beta }} - 3{{\hat d}_\alpha }{{\hat d}_\beta }.
\end{split}\eeq
The information of the initial state $\left|\Phi \right\rangle $ and the final state $\left| \Omega \right\rangle $ is encoded in the expected values of each dipole moment. We note that in our approximate calculation, when the entanglement phase $\delta \phi$ is in the 1st and 4nd superposition branches on the right side of Eq. \eqref{1}, the photon emission rate is independent of the entanglement phase. $d$ represents the distance between particle pairs and $\hat{d}$ represents unit vector of the relative position between particles. Repeated indices represent summation.

In following we take both the two magnetic dipole moments are spin magnetic dipole moment of an electron for example to clarify the details of  photon emission rate of entangled state. Now the magnetic dipole moment operator is:
\beq\label{7}
{{\vec \hat u}_e} =  - {g_e}{u_B}\frac{{\vec \hat s}}{\hbar } =  - {g_e}{u_B}\frac{{\vec \hat \sigma }}{2}.
\eeq
Here we have assumed that the particle spin is $\frac{1}{2}$ and $\vec \hat \sigma$ is pauli matrix. ${u_B}$ is Bohr magneton and ${g_e}$ is $Land\acute{e}\;g{\rm{ - }}factor$. Combining all relevant specific expressions Eq.  \eqref{1}\eqref{5}\eqref{20}\eqref{21}\eqref{7}, the relationship between $R$ and entanglement phase is:
\begin{widetext}
\beq\label{15}
R\left( {\phi ,k,d} \right) = \frac{{{u_0}{g_e}^2{u_B}^2{k^3}}}{{32\pi \hbar }}\left( {\frac{4}{3} + \frac{{\cos \left( \phi  \right)\left( {\sin \left( {kd} \right) + kd\left( {kd\sin \left( {kd} \right) - \cos \left( {kd} \right)} \right)} \right)}}{{2{k^3}{d^3}}}} \right).
\eeq
\end{widetext}
When $d$ approaches its limit values, $R$ is given by:
\begin{subequations}\label{16}
	\begin{align}
		&{R_{d \to 0}} = \frac{{{u_0}{g_e}^2{u_B}^2{k^3}}}{{48\pi \hbar }}\left( {2 + \cos \left( \phi  \right)} \right) \\
		&{R_{d \to \infty }} = \frac{{{u_0}{g_e}^2{u_B}^2{k^3}}}{{24\pi \hbar }}\label{16b}
	\end{align}
\end{subequations}

\begin{figure}
	\centering
	\subcaptionbox{The photon emission rate $R$ varies with entanglement phase $\delta \phi $ with $kd$ fixed.\label{f3a}}
	{\includegraphics[scale=0.435]{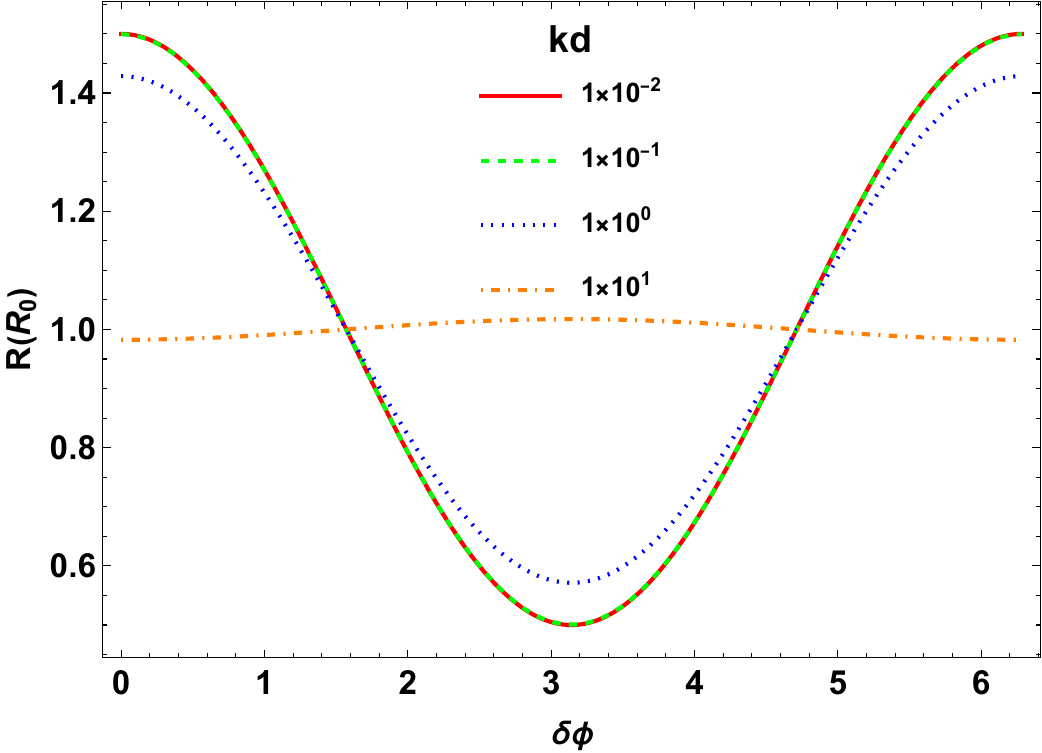}}
	\subcaptionbox{The photon emission rate $R$ varies with particle spacing $kd$ with $\delta \phi$ fixed.\label{f3b}}
	{\includegraphics[scale=0.45]{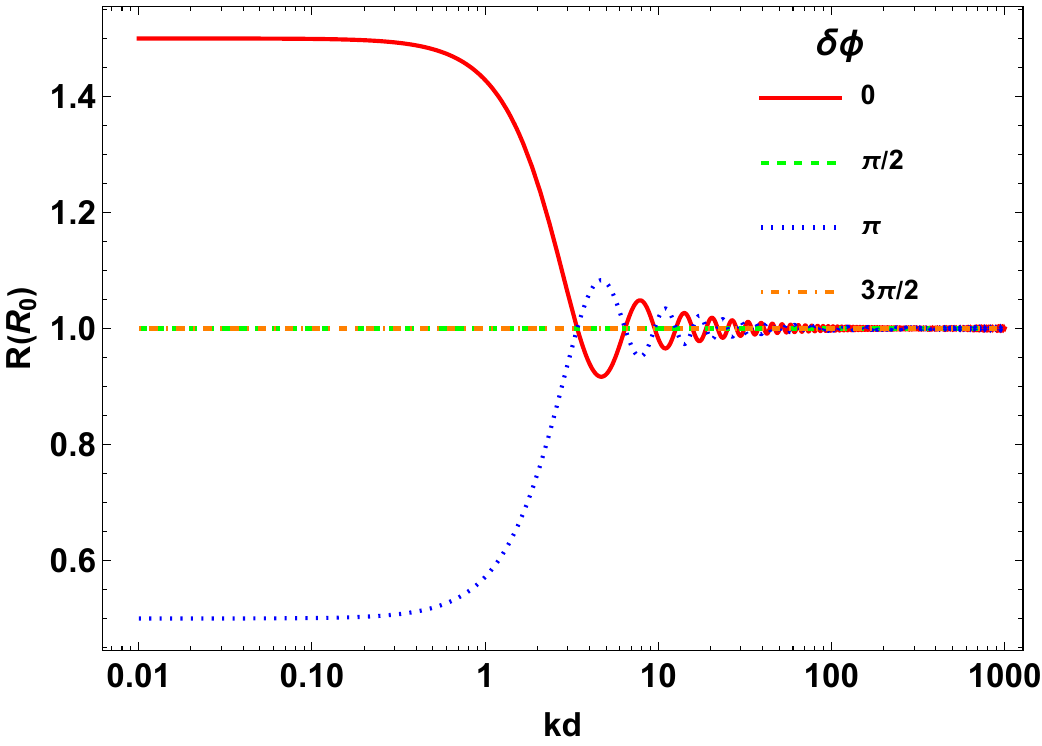}}
	\caption{Schematic diagram of particle pair transitions. In the numerical calculation, we set the particle spin quantum number to \(\frac{1}{2}\). $R$ stands for photon emission rate measured in ${R_0}$, $kd$ for spatial distance between two particles multiplied by the photon wave number $k$, $\delta \phi $ for entanglement phase.}\label{f3}
\end{figure}

Fig. \ref{f3} illustrates how the rate $R$ changes with variations in the induced entanglement phase and the separation distance between the two particles. We have measured the transition rate $R$ in units of ${R_{d \to \infty }}$ and denote it as ${R_0}$.

In Fig. \ref{f3a}, when $d$ is fixed at a small value, the  photon emission rate $R$ decreases with increasing entanglement degree within one half period of the entanglement phase. But for large $d$, $R$ will increase slightly within first half cycle. Moreover, as $d$ increases, the fluctuation of $R$ decreases. Eventually, when $d$ reaches a certain large value, it will no longer have any influence on $R$. In Fig. \ref{f3b}, when the entangled state has different entanglement phases, the variation of $R$ with respect to $d$ exhibits distinct characteristics. Entanglement can either inhibit or promote the transition rate, the greater the degree of entanglement, the stronger its influence—whether suppressing or enhancing $R$. The line with phase $\pi /2$ and  $3\pi /2$ coincide because they are equivalent in phase. When the distance between two particles approaches either infinitely close or infinitely far apart, $R$ tends to a stable value as given by Eq. \eqref{16}. However, when the separation between the particles becomes comparable to the wavelength of the emitted photons, $R$ exhibits fluctuations. We observe an interesting phenomenon where, when $kd$ takes certain values, $R$ has nothing to do with the entanglement phase, and all curves will intersect at the same point.  The blue line with entanglement phase $\pi$, the maximum entanglement state, reaches a state with the lowest transition rate and the most stable state as $d$ approaches zero.
   
\begin{figure}\centering
	\includegraphics[scale=0.47]{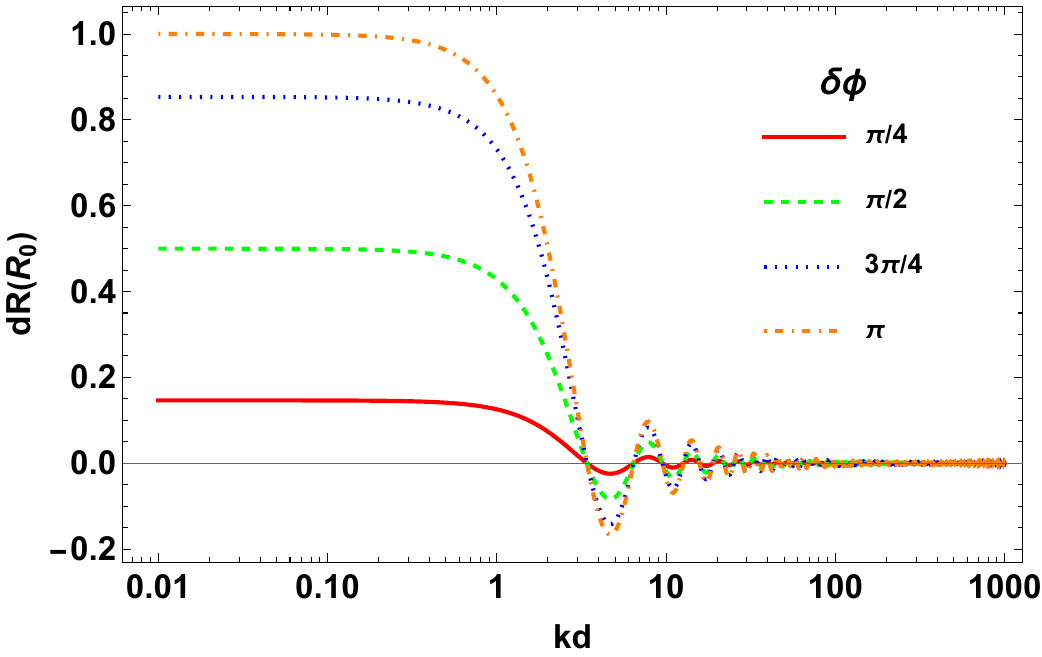}
	\caption{The photon emission rate difference of each entangled state as a function of particle distance. The horizontal axis, $kd$, is the distance between particle pairs multiplied by the photon wave number. The vertical axis, transition rate difference $dR$, is calculated as: ${\rm{dR = }}{{\rm{R}}_{\delta \phi  = 0}} - {{\rm{R}}_{\delta \phi }} $.}\label{f5}
\end{figure}

In order to clearly demonstrate the influence of entanglement phase formation on the photon emission rate, the most important thing is to compare the difference between the photon emission rates in cases where the entanglement phase is zero and nonzero. Fig. \ref{f5} shows that when the particle pairs approach each other and $d$ gradually decreases, the transition rate difference $dR$ first fluctuates up and down. As $d$ continues to decrease, $dR$ rises rapidly. When $d$ is less than $\frac{1}{k}$, the $dR$ gradually tends to a stable maximum value. This means that for all entangled states, the change in  photon emission rate in the smaller $d$ region is greater than the change in the larger $d$ region. Besides, when $d$ is fixed, the larger the degree of entanglement, the greater the change in emission rate. Clearly, the curve corresponding to the entanglement phase to $\pi$ is always at the outermost. 

\begin{figure}\centering
	\includegraphics[scale=0.44]{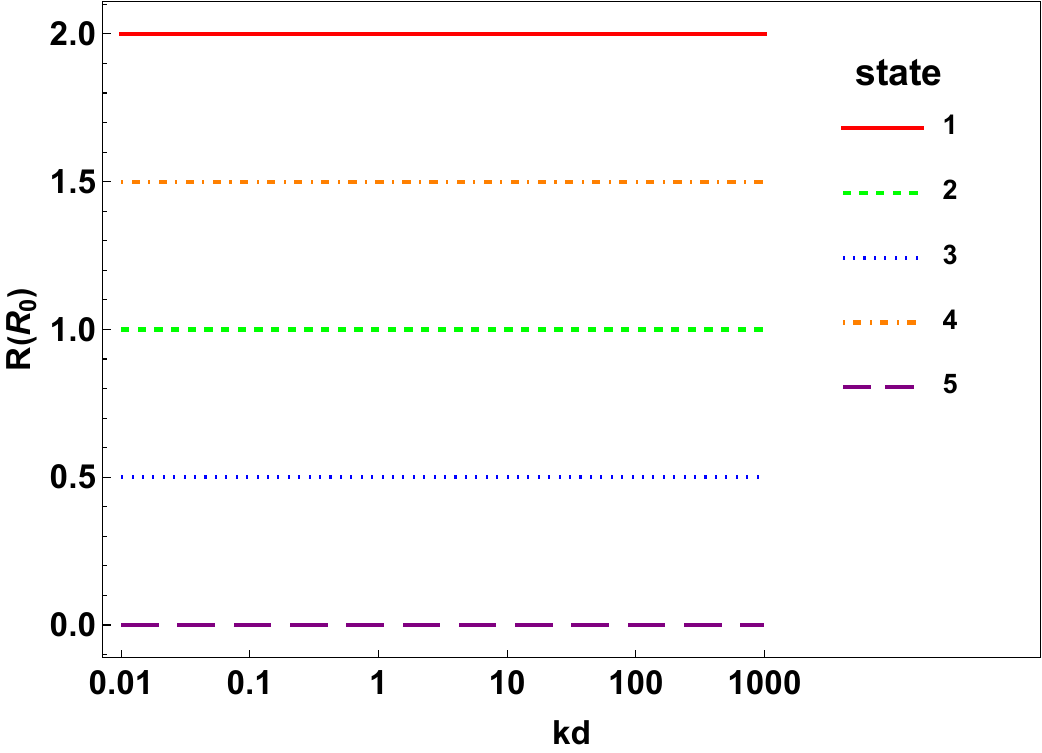}
	\caption{The photon emission rates of five non-entangled states: (1) $\left| { \uparrow  \uparrow } \right\rangle $, (2) $\left| { \uparrow  \downarrow } \right\rangle$, (3) $\frac{1}{{\sqrt 2 }}\left( {\left| { \uparrow  \downarrow } \right\rangle  + \left| { \downarrow  \downarrow } \right\rangle } \right)$, (4) $\frac{1}{{\sqrt 2 }}\left( {\left| { \uparrow  \uparrow } \right\rangle  + \left| { \downarrow  \uparrow } \right\rangle } \right)$, (5) ${\left| { \downarrow  \downarrow } \right\rangle }$. $R$ stands for photon emission rate measured in ${R_0}$, $kd$ for spatial distance between two particles multiplied by the photon wave number $k$.}\label{f6}
\end{figure}

It follows that when $ kd $ is fixed (excluding intersection points), the entanglement degree of the entangled state determined by the single parameter $\delta \phi$, as expressed in Eq. \eqref{1}, exhibits a one-to-one correspondence with the emission rate $R$. When $kd < 1$, the rate variation is maximized, making it most favorable for observation and comparison. As a result, we could select the photon emission rate, $R$, as our observable to detect and even measure the entanglement between the particles. However, there exists the possibility that other non-entangled states correspond to the same photon emission rate $R$ as a given entangled state. As an illustration, we present the emission rates of five product states in Fig. \ref{f6}. As shown in Fig. \ref{f6}, transition rates of  product states intersect with the entangled states in Fig. \ref{f3a} at certain values of $d$. But to distinguish these product states from the entangled state in Eq. \eqref{1}, comparing two emission rates at two values of $d$ is sufficient. So increasing the number of parameter $d$ values aids in the entangled states detection. However, in practical scenarios, there may exist all kinds of entangled states and even mixed states, complicating the factors affecting photon emission rates. Thus, based on our current calculations, we cannot confirm whether there exists a parameter space where the photon emission rates correspond to a set of entangled states.  

Although we fail to evaluate or measure entanglement from the perspective of quantum dynamical evolution \cite{Guhne:2008qic,Bernien:2017ubn,Huang:2020tih}, the photon emission rate $R$ at least may assist in our assessment of entanglement if we only restrict ourselves to the quantum state denoted by Eq. \eqref{1}.

\section{Conclusions}
In the PRL \cite{r3}, the analysis of the entanglement final state \eqref{1} was restricted to its entanglement detection via the entanglement witness \eqref{17}. Other properties of the entanglement final state were not taken into consideration. In this paper, we examine the effect of entanglement on the spontaneous photon emission rate of the distinguishable entanglement particle pair. We find that both the entanglement phase and the distance between particles significantly influence the  photon emission rate, assuming no external factors are present. When the two particles are close to each other, the more entanglement phase approaches $\pi$ within the range of $0$ to $\pi$, the more suppressive effect on the transition rate. Furthermore, when the spatial separation between the two particles is comparable to the wavelength of the emitted photon, the transition rate exhibits fluctuations. However, as the separation tends towards extremely large, the transition rate $R$ converges to a correspondingly stable asymptotic value independent of entanglement phase.

Although the entanglement between particles affects the evolution of quantum states, there is no one-to-one correspondence between the photon emission rate and entanglement degree so we cannot detect or measure entanglement by the emission rate. Only when considering specific quantum states \eqref{1} and fixing the parameter $d$ does the photon emission rate exhibit a one-to-one correspondence with the entanglement phase.  Entangled states do not involve changes in energy levels, thus lacking many observable effects. The alteration of photon emission rates is one of the few observable effects, which also reflects the non-locality of quantum mechanics. Thus, at least for state \eqref{1}, this finding provides a new perspective on detecting and even measuring entanglement through transition dynamics. In astronomy, there is growing interest in detecting quantum entanglement phenomena in the universe \cite{Chen:2017cgw,r30,Zhang:2023bka}. The photon emission rate may serve as a potential indicator for the astronomical detection of entanglement. Besides, the suppression of transitions by entanglement can be utilized to enhance the stability of similar quantum states in experimental setups. It provides insights and guidance for various practical applications, including quantum computation and communication \cite{r10,r11}, quantum sensing and detection \cite{Giovannetti:2004cas}, cold atom physics experiments \cite{Bloch:2008zzb}, and noise compensation in quantum networks \cite{Dur:1998qv}, etc. Moreover, it is reasonable to consider that the transition rate of such entangled particle pairs can be influenced by the background spacetime, which paves the way for probing curved spacetime quantum fields \cite{r4,r15,r16}.

In this paper, we only investigated the relationship between quantum entanglement and transition rates in a specific scenario. The possibility of detecting and even measuring quantum entanglement through quantum evolution under different models deserves more exploration in the future. Besides, in this specific derivations we have assumed only first-order perturbations play a role in particle pair transition. But the derivation can be easily extended to the higher order approximations in which more photons are emitted.  And the number of particles involved in the entanglement state is a crucial factor that affects the photon emission rates. The relationship between transition rate and entanglement in general many-body quantum systems with more particles remains unclear. Moreover, the transition rates of the QGEM entanglement state \eqref{1} with boundary condition and Unruh effect deserve further investigation.

\end{document}